\documentclass{elsart}
\usepackage{graphicx,amssymb}
\usepackage{float}
\usepackage{amsmath}
\usepackage{soul}
\usepackage{adjustbox}
\usepackage{multirow}
\usepackage[square,sort,comma,numbers]{natbib}
\usepackage{graphicx}
\usepackage{booktabs}
\usepackage{xcolor}
\usepackage{ulem}

\makeatletter
\def\elsartstyle{
\def\normalsize{\@setfontsize\normalsize\@xiipt{14.5}}
\def\small{\@setfontsize\small\@xipt{13.6}}
\let\footnotesize=\small
\def\large{\@setfontsize\large\@xivpt{18}}

\def\Large{\@setfontsize\Large\@xviipt{22}}
\skip\@mpfootins = 18\p@ \@plus 2\p@
\normalsize}
\makeatother

\def\url#1{{\ttfamily\def\/{/\discretionary{}{}{}}#1}}

\def\aj#1{{AJ}}%
\begin{document}

\begin{frontmatter}
\title{Impact of magnetic field on the gas mass fraction of galaxy clusters}
\author{Sandhya Jagannathan$^{a}$\thanksref{1}},
\author{Sunil Malik$^{a}$\thanksref{2}},
\author{Deepak Jain$^{b}$\thanksref{3}},
\author{T. R. Seshadri$^{a}$\thanksref{4}}
\address[a]{Department of Physics and Astrophysics, University of Delhi, Delhi 110007, India}
\address[b]{Deen Dayal Upadhyaya College, University of Delhi, Dwarka, New Delhi 110078, India}
\thanks[1]{E-mail: sjagannathan@physics.du.ac.in}
\thanks[2]{E-mail: sunil@physics.du.ac.in}
\thanks[3]{E-mail: djain@ddu.du.ac.in}
\thanks[4]{E-mail: trs@physics.du.ac.in}
\begin{abstract}
Magnetic fields have been observed in galaxy clusters with strengths of the order of $\sim \mu$G. The non-thermal pressure exerted by magnetic fields also contributes to the total pressure in galaxy clusters and can in turn affect the estimates of the gas mass fraction, $f_{gas}$. In this paper, we have considered a central magnetic field strength of $5\mu$G, motivated by observations and simulations of galaxy clusters. The profile of the magnetic field has also been taken from the results obtained from simulations and observations. The role of magnetic field has been taken into account in inferring the gas density distribution through the hydrostatic equilibrium condition (HSE) by including the magnetic pressure. We have found that the resultant gas mass fraction is smaller with magnetic field as compared to that without magnetic field. However, this decrease is dependent on the strength and the profile of the magnetic field. 
We have also determined the total mass using the NFW profile to check for the dependency of $f_{gas}$ estimates on total mass estimators. From our analysis, we conclude that for the magnetic field strength that galaxy clusters seem to possess, the non-thermal pressure from magnetic fields has an impact of $\approx 1~\%$ on the gas mass fraction of galaxy clusters. However, with upcoming facilities like Square Kilometre Array (SKA), it can be further expected to improve with more precise observations of the magnetic field strength and profile in galaxy clusters, particularly in the interior region.
\end{abstract}

\end{frontmatter}

\section{Introduction}
\label{intro}

Magnetic fields have been observed at various length scales from stars to galaxies and galaxy clusters~\citep{clarke2004,go2006,nv2010,akahori2011,reiss2014,ber2016,va2016}.
These fields have a bearing on different physical processes such as star formation \citep{sch2009}, confinement of cosmic-rays in the galactic arms and inside the core of the galaxy clusters~\citep{kush2009} among others. It has been suggested that these fields can also have an influence on the formation of large-scale structures such as galaxy clusters and in their virialization~\citep{shibu2014}. There are several indirect methods to probe the magnetic field strength and its structure in galaxies and galaxy clusters such as synchrotron radiation and Faraday rotation of polarised radiation from radio sources present inside or in the background of these structures~\citep{tay1993,car2002,go2004,clarke2004,fer2012,sunil2017}.
Using Faraday rotation observations, it had been inferred that galaxy clusters have central magnetic fields of the order of $\sim10\mu$G with a coherence length of $10-20$ kpc~\citep{dre1987,fer1999,tay2001,allen2001,tay2001,eilek2002,clarke2004,va2016,vacca2012}.
Further, it is also well established that the strength of these fields varies with the dynamical stage of the clusters, being more for relaxed clusters compared to the unrelaxed ones. However, recent simulations of magnetic field in galaxy clusters suggest that the central magnetic field strength ranges from $3-5\mu$G~\citep{vazz,domi}. The associated uncertainties in the central magnetic field strength are further expected to improve with future radio surveys such as the Square Kilometre Array~(SKA), which will provide precise information about the characteristics of magnetic fields over a range of scales~\citep{jh2015}.
 \par
The non-thermal pressure exerted by magnetic field has been investigated in several previous studies to gauge its effect on the total mass and the distribution of gas density~\citep{loeb,koch,lagana,Gopal_2010}. 
For instance, Lagan\'a et al.~(2010)~\citep{lagana} 
incorporated the pressure due to magnetic field, cosmic rays, and turbulence in the hydrostatic equilibrium condition to compute the change in the total mass of the clusters. In their analysis, they have observed that the inclusion of non-thermal pressure leads to an increase of about 10 to 35 $\%$ in the total mass of the cluster. This was further used to explain the inconsistency found in the total mass inferred from different measurements such as those from X-ray and weak lensing~\citep{wu1998,wu2000}. 
However, there has been another approach suggested wherein the magnetic field directly affects the rearrangement of the gas density keeping the total mass and the temperature profile unaffected. In previous studies, Koch~et~al.~(2003)~\citep{koch} 
and Gopal $\&$ Roychowdhury~(2010)~\citep{Gopal_2010}
have followed this route to deduce the change in the gas density profiles by assuming the dependence of magnetic field on the gas density as $B(r) \propto (\rho_{g}(r))^{\gamma}$, where $\gamma$ is the shape parameter~\citep{cg2007,dolag}. 
They have noted that the inclusion of magnetic field reduces the gas density (e.g., see Section $3$ of Koch~et~al.~(2003)~\citep{koch}) 
and found that the impact of magnetic field was larger in the core as compared to that in the outer regions of the cluster.  

In this paper, we aim to investigate the effect of magnetic field on the gas mass fraction, $f_{gas}$, of a sample of galaxy clusters. The gas mass fraction is defined as the ratio of the gas mass to the total mass of the cluster, $f_{gas}(r) = M_{g}(<r)/M_{total}(<r)$~\citep{sazarin1986,sasaki,evrad1997,ettori2003,ettori2009}.
 It serves as an alternate probe to constrain cosmological parameters since this ratio is considered to be indicative of the baryon fraction, $\Omega_{b0}/\Omega_{m0}$, in the universe where $\Omega_{b0}$ and $\Omega_{m0}$ are the baryon density and the total matter density of the universe today, respectively~\citep{White,sasaki,allen2003,allen2004,allen2008,ettori2009,allen2011}.
By taking advantage of the availability of number density profiles and their corresponding best fit parameters for 35 galaxy clusters in LaRoque~et~al.~(2006)~\citep{laro}, we aim to extend the study conducted by Koch et. al. (2003)~\citep{koch} and infer the resultant change in the gas mass fraction through the effect of magnetic field on the gas density profile. 
In addition, to gauge differences in the gas mass fraction estimates using different total mass estimators, we have also employed Navarro, Frenk and White (NFW) profile~\citep{NFW} to compute the total mass of the cluster. We have also evaluated the change in the gas mass fraction for our sample as a function of $B_{0}$ and $\gamma$. In our analysis, we have considered the following cosmological parameters: $H_{0} = 70 ~\text{km}~ \text{s}^{-1}~ \text{Mpc}^{-1}, \Omega_{m0} = 0.30$ and $\Omega_{\Lambda0} = 0.70$.
\par
This paper is organized as follows: Section \ref{MF} details the estimation of gas mass, total mass and the gas mass fraction for our sample, with and without magnetic field.  
In the same section, we also discuss the effect of employing the NFW profile in estimating the total mass. Further in Section \ref{met}, we explain the methodology adopted for the analysis. In Section \ref{rs}, the results of our analysis our provided and in the subsequent section, we conclude with our major results.

\section{Estimation of gas mass, total mass and the gas mass fraction} \label{MF}
The intracluster medium comprises of magnetized plasma at a temperature of $\sim 10^8$~K~\citep{sazarin1986,go2004,ettori2013}. It predominantly emits in X-ray, which is often used in the estimation of gas density and other physical parameters of the cluster~\citep{sazarin1986,go2004,laro,bona2006,landry,ettori2013}. 
For our analysis, we have considered a sample of $35$ galaxy clusters distributed in the redshift range, 0.14 to 0.89, from LaRoque et al.~(2006) ~\citep{laro}. They have modelled the galaxy clusters using an isothermal-$\beta$ profile with the central $100$~kpc removed from the analysis. This was done to remove the ambiguity from the core of relaxed clusters which show excess X-ray in the central region. Hence, this profile fits both the relaxed and unrelaxed clusters equally well. Koch et al. (2003) also considered an isothermal-$\beta$ profile to explain the role of magnetic fields. Since the work reported in the present paper is an extension of their work, we consider the same profile. 
The expression for the number density as given by the isothermal-$\beta$ profile is,
\begin{equation}
n_{e}(r) =  n_{0}\left( \frac{1}{1+(r/r_{c})^{2}}\right)^{3\beta/2} \label{1}
\end{equation}
The parameters $n_{0}$ and $r_{c} $ represent the central number density and the core radius, respectively. 
Using Eq.~\ref{1} and temperature, $T_{0}$, we compute the gas density, total mass and the gas mass fraction at the radii, $r_{2500,HSE}$ and $r_{\tiny{500,HSE}}$, which are defined as the radii at which the total density of the cluster is $2500$ and $500$ times the critical density of the universe, $\rho_{c}$, at the cluster's redshift, respectively. Further, substituting for the number density profile from Eq. \ref{1}, the gas mass within a radius, $r$, turns out to be,
\begin{align}
M_{\tiny{g,HSE}} (<r) &= 4\pi  \mu_{e} m_{p} \int_{0}^{r} {\tilde{r}}^{2} n_{e}(\tilde{r}) d\tilde{r} \\
					&= 4 \pi \mu_{e}m_{p}\int_{0}^{r} \left( n_{0} \left ( \frac{1}{1+(\tilde{r}/r_{c})^{2}} \right )^{3\beta/2}   \right) \tilde{r}^{2} d\tilde{r}, \label{3}
\end{align}
where $m_p$ and $\mu_{e}$ are the mass of the proton and the mean molecular weight of the electron, respectively. We have considered the value of $\mu_{e}$ to be 1.624~\citep{vk}. Using the hydrostatic equilibrium condition, the total mass is given as (e.g., Eq. 2.12 in~\citep{Gopal_2010}),
 
\begin{equation}
M_{\tiny{HSE}}(r) = \frac{-kT(r)r}{G\mu m_{p}}\left(\frac{d\ln{n_{e}}(r)}{d\ln r} + \frac{d\ln{T(r)}}{d\ln r}\right). \label{4} \\
\end{equation}
Here, $\mu$,~$T(r)$,~$k$, and $n_{e}(r)$ represent the mean molecular weight, temperature, Boltzmann constant and the number density profile of the electrons, respectively. 
Using the gas mass estimated from Eq.~\ref{3} and the total mass estimated from Eq.~\ref{4} (Hydrostatic equilibrium condition), the gas mass fraction, henceforth denoted as $f_{\tiny{gas,HSE}}$, can be expressed as,
\begin{align}
f_{\tiny{gas,HSE}}(<r) =\frac{ M_{\tiny{g,HSE}}(<r)}{M_{\tiny{HSE}}(<r)}. \label{5}
\end{align}

\subsection{Effect of magnetic field on $M_{\tiny{g,HSE}}$ and $f_{gas,HSE}$} \label{MF1}

The high magnetic field strengths found in galaxy clusters 
could play a non-trivial role in the dynamics of the gas by its direct impact on the distribution of electron number density through its contribution to the total pressure~\citep{koch,Gopal_2010}.
As discussed in Section \ref{intro}, there are two approaches suggested to incorporate this effect into the cluster dynamics and we follow the approach discussed by Koch~et~al.~(2003)~\citep{koch} and Gopal $\&$ Roychowdhury~(2010)~\citep{Gopal_2010}. In this approach, inclusion of magnetic field in the hydrosatic equilibrium condition modifies the gas density, keeping the total mass unaltered. It can be justified by the fact that the total mass of the cluster is dominated by the dark matter mass (with the gas and the baryonic content constituting $\sim 15\%$ of the total mass) and that magnetic field directly interacts only with baryons and not dark matter~\citep{sazarin1986}. 
We also check for the validity of the assumption that the total mass remains unaltered by the inclusion of magnetic field, i.e., $M_{\tiny{HSE}} \approx$ $M_{\tiny{HSE,B}}$, in Section \ref{met}. 
The hydrostatic equilibrium condition in the presence of magnetic field is given by (e.g., Eq.~5 of~\citep{koch}),
\begin{align}
M_{\tiny{HSE,B}}(r) &= \frac{-kT_{\tiny{B}}(r)r}{G\mu m_{p}}\left(\frac{d\ln{n_{\tiny{e,B}}}(r)}{d\ln r} + \frac{d\ln{T_{\tiny{B}}(r)}}{d\ln r}\right) -\frac{r^{2}}{G\rho_{\tiny{g,B}}(r)}\frac{dP_{\tiny{B}}(r)}{dr}. \label{6}
\end{align}
Here, $T_{B}(r)$ denotes the temperature of the gas in the presence of magnetic field. The quantities $n_{e,B}(r)$,
$\rho_{g,B}(r)$, and $P_{B}(r)$ denote the number density of electrons in the presence of magnetic field, gas density in the presence of magnetic field and the additional pressure that arises due to magnetic field, respectively. In earlier studies, magnetic field observations of galaxy clusters
were interpreted assuming a constant magnetic field distribution~\citep{car2002,go2004,bo2016}. Later, it was suggested that the magnetic field distribution follows the gas density distribution~\citep{jaffe1980,brunetti2001}. Dolag et 
al.~(2001)~\citep{dolag} found a correlation between the r.m.s of the rotation measure and the X-ray
emission in galaxy clusters using results from MHD simulations and observations. The magnetic field distribution was modelled using the form,
\begin{equation}\label{7}
B(r) = B_{0}\left(\frac{\rho_{\tiny{g}}(r)}{\rho_{\tiny{g0}}}\right)^{\gamma}.
\end{equation}
Here, $\rho_{g0}$ and $B_{0}$ represent the gas density and the magnetic field at the center of the cluster, respectively and $\gamma$ denotes the shape parameter. The same form was also later used by Koch~et~al.~(2003)~\citep{koch} to study the impact of magnetic fields on the Sunyaev-Zel'dovich effect in galaxy clusters. It was also used by Colafrancesco $\&$ Giordano~(2007)~\citep{cg2007} and Lagan\'a et al.~(2010)~\citep{lagana} to study the effect of non-thermal pressure in their respective studies. 
The value of the shape parameter, $\gamma$, depends on the nature of the cluster. For example, during cluster formation, the flux-freezing condition results in $\gamma$ of the order of $2/3$~\citep{dolag,lagana}. Similarly, in the case of equipartition of energy in galaxy clusters, it can be approximated as $0.5$. Using the results from simulations and observations of galaxy clusters, Dolag et al.~(2001)~\citep{dolag} reported that the shape parameter for unrelaxed clusters such as A119 is of the order of $\sim0.9$. In our analysis, we have considered the two extreme values of $\gamma$ namely, $0.5$ and $0.9$ for our sample of galaxy clusters. Lagan\'a et al.~(2010)~\citep{lagana} also considered $\gamma$ in the range $0.5$ to $0.9$ in their analysis, which is consistent with our assumptions. 
Assuming that the temperature is constant and is not affected by magnetic field ($T(r) = T_{B}(r) = T_{0}$), the expression for $\rho_{g,B}(r)$ inferred from equating Eq.~\ref{4} and \ref{6} turns out to be,~
\begin{align}\label{8}
\rho_{\tiny{g,B}} (r) &= \rho_{g}(r) \exp \left({\frac{{B_{0}}^{2}}{2\mu_{0}}\frac{\mu m_{p}}{{\rho_{B_{0}}}^{2\gamma}kT_{0}} \int_{r}^{r_{l}} \frac{({\rho_{\tiny{g,B}}(\tilde{r})}^{2\gamma})^{'}}{\rho_{\tiny{g,B}}(\tilde{r})} d\tilde{r}}\right). 
\end{align}
Here, $\rho_{B_{0}}$ denotes the central gas density in the presence of magnetic field. The above expression is solved iteratively for the values of $B_{0}$ and $\gamma$ considered in our study by first substituting $\rho_{g}(r)$ in the place of $\rho_{g,B}(r)$ on the right hand side. We consider the expression obtained after the first iteration for our analysis . In Eq.~\ref{8}, $r_{l}$ denotes the radius where $\rho_{g,B}(r_{l}) = \rho_{g}(r_{l})$. Similar to the analysis done in the previous section, we can describe $M_{\tiny{g,HSE,B}}$ and $f_{\tiny{gas,HSE,B}}$ using the expression for the gas density and the total mass, $\rho_{\tiny{g,B}}$ and $M_{\tiny{HSE}}$, given in Eq.~\ref{8} and Eq.~\ref{4}, respectively. The gas mass, $M_{\tiny{g,HSE,B}} $ and the gas mass fraction, $f_{\tiny{gas,HSE,B}}$, are given as follows,
\begin{align} 
M_{\tiny{g,HSE,B}}(<r) &= 4 \pi \mu_{e} m_{p} \int_{0}^{r} {\tilde{r}}^{2} \rho_{\tiny{g,B}} (\tilde{r}) d\tilde{r} \label{10}\\
f_{\tiny{gas,HSE,B}}(<r) &=  \frac{M_{\tiny{g,HSE,B}}(<r)}{M_{\tiny{HSE}}(<r)}. \label{11}
\end{align}

\subsection{Estimating the total mass using the NFW profile} \label{MF2}

In the literature, the total mass of a galaxy cluster is also modelled using the NFW profile \citep{NFW}, which was derived by modelling the dark matter distribution in N-body simulations. It is representative of the total mass of the cluster as the dark matter dominates the total mass content. The NFW density profile is given by,

\begin{equation}
\rho_{\tiny{NFW}}(r) = \frac{\rho_{c} \delta_{c}}{(r/r_{s})(1+r/r_{s})^{2}}. \label{12}
\end{equation}
Here, $\delta_{c}$ refers to the characteristic overdensity of the halo. The radial scale $r_{s}$ is equal to $r_{200}/c$, where $r_{200}$ and $c$ are the virial radius  and the concentration parameter of the cluster, respectively. Since dark matter dominates the total mass of the galaxy cluster, the latter can be approximated from the NFW profile,

\begin{align}
M_{\tiny{NFW}} (r) & = \int_{0}^{r} 4 \pi \tilde{r}^{2} \rho_{NFW}(\tilde{r}) d\tilde{r} \label{13} \\ 
                  &= \int_{0}^{r} 4 \pi \tilde{r}^{2} \left(\frac{ \rho_{c} \delta_{c}}{(\tilde{r}/r_{s})(1+\tilde{r}/r_{s})^{2}}\right)d\tilde{r} \label{14}.
\end{align} 
With the change in the total density and the total mass due to the use of the NFW profile, the estimated values of $r_{\tiny{2500,HSE}}$ and $r_{\tiny{500,HSE}}$ also change. We denote these new quantities as $r_{\tiny{2500,NFW}}$ and $r_{\tiny{500,NFW}}$. As a result, the estimated gas mass at these radii is also modified. The gas mass fraction ($f_{\tiny{gas,NFW}}$) is now given by $M_{\tiny{g,NFW}}/M_{\tiny{NFW}}$. Similarly for a non-zero magnetic field, the total mass maybe assumed to be unaffected but that the gas mass changes to $M_{\tiny{g,NFW,B}}$. Hence, the new gas mass fraction in the presence of magnetic field becomes $f_{\tiny{gas,NFW,B}} = M_{\tiny{g,NFW,B}}/M_{\tiny{NFW}}$. The sample for the NFW mass profile was obtained from LaRoque~et~al.~(2006)~\citep{laro}. 

In the subsequent sections, we will compute the gas mass fractions for our sample as discussed above. We expect the gas mass fraction to decrease with the increase in the strength of magnetic field~(e.g., see Fig.~\ref{img3}). The change also depends on the value of the shape parameter $\gamma$ and the radius at which the quantity is determined. 

\section{Methodology} \label{met}

As discussed in Section \ref{MF2}, the mass calculated using the hydrostatic equilibrium condition is assumed to be the same with or without magnetic field. In order to check how good this assumption is, we calculated the variation, $\delta_{M} = [M_{\tiny{HSE,B}}(r)-M_{\tiny{HSE}}(r)]/M_{\tiny{HSE}}(r)\times100 \%$ for the galaxy clusters in our sample. For $B_{0}=5\mu$G and $\gamma=0.9$, clusters for which the variation was more than $\approx 1 \%$ were excluded as outliers at both $r_{2500,HSE}$ and $r_{500,HSE}$ using the Chauvenet's criterion\footnote{According to the Chauvenet's criterion, an observation $x_{i}$ is an outlier if the following expression holds true: $erfc\left(\frac{\mid{x_{i}-\bar{x}}\mid}{\sigma_{i}}\right) < \frac{1}{2N}$, where $x_{i}$ is the observation in question, $\bar{{x}}$ is the average of the sample, $\sigma_{i}$ is the standard deviation of the observation, and N is the total number of observations.}. The values for the rest of the cases are listed in Table~\ref{table0}.
After the removal of the outliers from the cluster sample, we calculated the average of the variation, $\overline{\delta_{M}}$ , for the remaining clusters. The values of $\overline{\delta_{M}}$ are listed in Table~\ref{table0} for different values of $B_{0}$, $\gamma$, and $r$ for our sample of clusters. Since we are interested in the cases where the magnetic field does not significantly alter the total mass, we further imposed $\overline{\delta_{M}}$ as a cut-off to exclude all those clusters for which the value of $\delta_{M}$ exceeded this average. The reduced sample is also provided in Table~\ref{table0}. We have also provided the analysis for $B_{0}=10~\mu$G for comparison purposes (not included in the final results) even though observations suggest that such high magnetic fields are unlikely.

\vspace{1cm}

\begin{table}[h]  
\renewcommand\arraystretch{1}
\large
\caption{The sample size and the cut-off values, $\overline{\delta_{M}}$, for our sample.}
\vspace{0.5cm}
\label{table0}
\begin{adjustbox}{width=1\textwidth}
\small
\begin{minipage}{20cm}
\begin{center}
\begin{tabular}{ | c | c| c c | c c| c c|} 
\hline
 Magnetic field~($B_{0}$) & $\gamma$ &  \multicolumn{2}{c|}{$\text{N}_{o}$\footnote{$N_{o}=$ Number of clusters after the removal of outliers.}} & \multicolumn{2}{c|}{$\overline{\delta_{M}}$ after removing outliers~($\%$)} & \multicolumn{2}{c|}{$\text{N}_{f}$\footnote{$N_{f}=$ Number of clusters in the final sample.}} \\ 
\hline
 & & $r_{2500,HSE}$ & $r_{500,HSE}$ & $r_{2500,HSE}$ & $r_{500,HSE}$ & $r_{2500,HSE}$ & $r_{500,HSE}$ \\
\cline{2-8}
\multirow{2}{*}{5$\mu$G} & $0.5$ &   34 & 34 & $0.67$ & $0.67$ & 21 & 21 \\
\cline{2-8}
                     &  $ 0.9$ & 32 & 33 & $0.35$ & $0.15$ & 21 & 24 \\
\cline{2-8} 
 \multirow{2}{*}{10$\mu$G\footnote{For comparison purposes only. Not included in the final results.}} & $0.5$ &   34 & 34 & $2.67$ & $2.67$ & 21 & 21 \\
\cline{2-8}
                     &  $ 0.9$ & 32 & 33 & $1.42$ & $0.59$ & 21 & 24 \\
 
                            
\hline
\end{tabular}
\end{center}
\end{minipage}
\end{adjustbox}
 \end{table}

 \section{Results and Analysis} \label{rs}

Using the reduced sample obtained in the previous section, we investigate the impact of magnetic field on the gas mass fraction for our sample of clusters corresponding to the cases of hydrostatic equilibrium as well as NFW. 
Along with that, to compare the values of $f_{gas}$ of our sample with the theoretical estimate of the gas mass fraction derived from the assumed cosmological parameters, we have removed the contribution of the stellar content, $f_{stars}$, from the theoretical value, where, $f_{stars}$ is estimated from the analysis done by Lagan\'a et al.~(2013)~\citep{lagana2013} by computing the average value of $f_{stars}$ for the $12$ clusters that overlapped with our sample of clusters left after the removal of outliers. The stellar fractions estimated at $r_{2500,HSE}$ and $r_{500,HSE}$ are $0.014 \pm 0.001$ and $0.015 \pm 0.001$, respectively. The same applies for the NFW profile. 
\par
\begin{figure}[h]
\centering
\includegraphics[width=0.8 \textwidth]{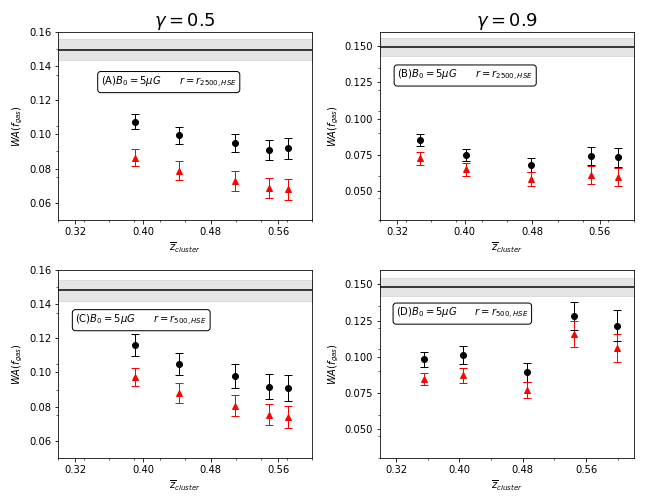}
\caption{\small{\textit{\emph{\bf{Left panel:}} The distribution of the weighted averages ($WA$) of $f_{\tiny{gas,HSE}}$ with (red solid triangle) and without (black solid disc) magnetic field with respect to the mean redshift in every redshift bin, $\bar{z}_{cluster}$ (see Section \ref{met} for more details). 
Here, black solid line denotes the theoretical value of $f_{gas}=$ $ \Omega_{b0}/\Omega_{m0} - f_{stars}$ estimated from the assumed cosmological parameters with the 1$\sigma$ region shaded in grey.  \emph{\bf{Right Panel:}} Same as the left panel but for the case of $\gamma=0.9$.}}}
\label{img1}
\end{figure}

After conducting our analysis, we found that the average percentage change in $f_{gas,HSE}$ for $B_{0} = 5\mu$G and $\gamma=0.9$ is $0.13~\%$ and $0.07~\%$ at $r_{2500,HSE}$ and $r_{500,HSE}$, respectively. The same increases to $0.60~\%$ and $0.40~\%$ at $r_{2500,HSE}$ and  $r_{500,HSE}$, respectively for $\gamma=0.5$. This is evident from the expression given in Eq.~\ref{7}. As the value of $\gamma$ decreases, the contribution of magnetic field at a particular radius increases. The behaviour of the weighted average of $f_{gas,HSE}$ and $f_{gas,HSE,B}$ is shown in Figure~\ref{img1}. 
\par
\begin{figure}[h]
\centering
\includegraphics[width=0.8 \textwidth]{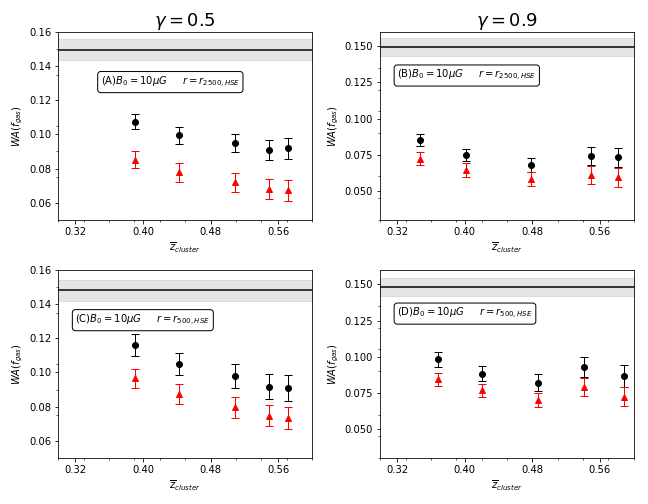}
\caption{\small{\textit{\emph{\bf{Left panel:}} The distribution of the weighted averages ($WA$) of $f_{\tiny{gas,HSE}}$ with (red solid triangle) and without (black solid disc) magnetic field with respect to the mean redshift in every redshift bin, $\bar{z}_{cluster}$ (see Section \ref{met} for more details). 
Here, black solid line denotes the theoretical value of $f_{gas}=$ $ \Omega_{b0}/\Omega_{m0} - f_{stars}$ estimated from the assumed cosmological parameters with the 1$\sigma$ region shaded in grey.  \emph{\bf{Right Panel:}} Same as the left panel but for the case of $\gamma=0.9$.}}}
\label{img2}
\end{figure}

Although, a central magnetic field strength of $10\mu$G in galaxy clusters is rare, we have considered the same to demonstrate the increase in the change in $f_{gas,HSE}$ with magnetic field.  For $B_{0}=10\mu$G, the resultant percentage change in $f_{gas,HSE}$ increases as shown in Figure~\ref{img2}, where the weighted average value of $f_{gas,HSE}$ and $f_{gas,HSE,B}$ are estimated as a function of the mean redshift in every redshift bin, $\bar{z}_{cluster}$. The average percentage change in $f_{gas,HSE}$ for $B_{0}=10 \mu$G and $\gamma=0.9$ is $0.54~\%$ and $0.31~\%$ at $r_{2500,HSE}$ and $r_{500,HSE}$, respectively. The values increase to $2.21~\%$ and $1.65\%$ at $r_{2500,HSE}$ and $r_{500,HSE}$, respectively for $\gamma=0.5$. 
\par
Further, this difference in the gas mass fraction was found to be affected by changing the total mass estimator. To elaborate on this possibility, we have also estimated $f_{gas,NFW}$ and $f_{gas,NFW,B}$ using the NFW profile. From this exercise, we have noted that the change in $\Delta f_{\tiny{gas,NFW}}$ is larger at radii, $r_{\tiny{2500,NFW}}$ as well as $r_{\tiny{500,NFW}}$ in the NFW analysis as compared to the case of HSE. The average percentage change in the value of $f_{gas,NFW}$ with and without magnetic field are listed in Table~\ref{table1}. 
\vspace{2cm}
\begin{table}[H]  
\centering
\renewcommand\arraystretch{1}
\large
\caption{\small The average percentage change in the value of $f_{gas}$ in the presence of magnetic field of different strengths and different values of $\gamma$ are listed at $r_{2500,HSE}$, $r_{2500,NFW}$ and $r_{\tiny{500,HSE}}$, $r_{500,NFW}$.} 
\vspace{0.5cm}
\label{table1}
\begin{adjustbox}{width=1\textwidth}
\small
\begin{minipage}{20cm}
\begin{tabular}{| c | c | c  c  c  c | c  c  c  c |} 
\hline
 Magnetic field~($B_{0}$) & $\gamma$  &  \multicolumn{4}{c|}{HSE}  & \multicolumn{4}{c|}{NFW} \\ 
\hline
 & & \multicolumn{4}{c|}{\footnote{$\Delta f_{gas,HSE}=$ $ \frac{1}{N}\Sigma_{i}^{N}\left(\frac{f_{gas,HSE_{i}}-f_{gas,HSE,B_{i}}}{f_{gas,HSE_{i}}} \times 100\right) ~ (\% ).$} $ \Delta f_{gas,HSE}$ ($\%$)} & \multicolumn{4}{c|} {\footnote{$\Delta f_{gas,NFW}=$ $\frac{1}{N}\Sigma_{i}^{N}\left(\frac{f_{gas,NFW_{i}}-f_{gas,NFW,B_{i}}}{f_{gas,NFW_{i}}} \times 100 \right) ~ (\% ).$} $ \Delta f_{gas,NFW}$($\%$)} \\ \cline{3-10}

  & & $r_{2500,HSE}$ & $N_{f}$ & $r_{\tiny{500,HSE}}$ & $N_{f}$ & $r_{2500,NFW}$  & $N_{f}$ & $r_{500,NFW}$  & $N_{f}$ \\\cline{3-10}
 \multirow{2}{*}{$5\mu$G} & 0.5 &  0.60   & 21   & 0.41  &  21 & 0.96  & 35 &  0.87  & 35 \\
\cline{2-10}
 & 0.9 & 0.13   & 21   & 0.07  &  24 & 0.34  & 35 & 0.27 & 35 \\
\cline{2-10}  
 \multirow{2}{*}{$10\mu$G\footnote{For comparison purposes only. Not included in the final results.}} & 0.5 &  2.21   & 21   & 1.65   &  21 & 3.79  & 35 &  3.43  & 35 \\
\cline{2-10}
 & 0.9 & 0.54   & 21   & 0.31  &  24 & 1.36  & 35 &  1.08  & 35 \\
\cline{2-10}                                
\hline
\end{tabular}
\end{minipage}
\end{adjustbox}
 \end{table}

\begin{figure}[h!]
\centering
\includegraphics[width=0.9 \textwidth]{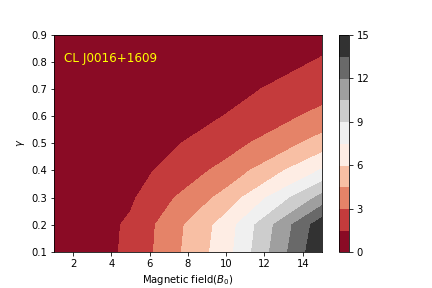}
\caption{\small{\textit{ The variation of the percentage change in the value of $f_{gas,HSE}$ at $r_{500,HSE}$ with the central value of magnetic field, $B_{0}$, and $\gamma$ for the cluster, $CLJ0016+1609$. The colorbar indicates the percentage change in $f_{gas,HSE}$.}}}
\label{img3}
\end{figure} 
\section{Discussions and Conclusions} \label{CD}
Observations suggest that the intracluster medium is magnetized, which leads to variations in the plasma processes on scales much smaller than the scale of the galaxy cluster.  These changes occurring on micro scales due to magnetic field can affect large scale properties such as thermal conduction and viscosity by making them anisotropic~\citep{mal2001}. This happens because the charged particles are guided along the magnetic field lines and their radius of gyration is much smaller than the Coulomb collisional mean free path. 
In addition, magnetic field also contributes to the total pressure over and above the thermal pressure along with other non-thermal sources like, turbulence and cosmic rays. In particular, the effect of magnetic field on the physical parameters of the clusters such as the total mass and density profile have been investigated in several earlier studies \citep{loeb,koch,lagana,Gopal_2010}. These can be classified into two approaches. In the first approach, the total mass is altered by the incorporation of magnetic field in the hydrostatic equilibrium condition while keeping the gas density unchanged. The second approach involves keeping the total mass constant and altering the gas density profile by the inclusion of magnetic field. \par
The study conducted by Lagan\'a et al.~(2010)~\citep{lagana} follows the first approach in which they have reported that the total mass varies upto $\sim35~\%$ by including contribution from magnetic field, cosmic rays, and turbulence. 
On the other hand, Koch~et~al.(2003)~\citep{koch} and Gopal $\&$ Roychowdhury (2010)~\citep{Gopal_2010} followed the second approach in their analysis. Koch~et~al.~(2003)~\citep{koch} considered one cluster and modelled its gas density using the $\beta$-model and concluded that the modified gas density profile is lowered by $10~\%$ to $20~\%$ in the innermost part of the cluster where the magnetic field is strongest. \par
In this paper, we have followed the second approach (similar to Koch~et~al.(2003) \\ \citep{koch}) and have quantified the effect of magnetic field on the gas mass fraction of a sample of galaxy clusters taken from LaRoque~et~al.~(2006) \citep{laro}.
We have extended the work done previously by Koch et al.~(2003) \citep{koch} and have incorporated the effect of magnetic field into $f_{gas}$. This approach has not been studied before in the context of gas mass fraction. We have also checked whether our assumption of the total mass remaining unchanged holds true by computing the variation for different combinations of $B_{0}$, $\gamma$, and $r$. 
Due to the above filters, although our sample size reduced, we were still left with at least 21 clusters for which we have evaluated the modified gas density profile by including the non-thermal pressure due to the magnetic field in the hydrostatic equilibrium condition. 
We have assumed a model of magnetic field dependence with radius in the galaxy cluster as $B(r) \propto \rho_{g}(r)^{\gamma}$ \citep{dolag,col1996,lagana} and we have considered two different values of $\gamma$ namely, $0.5$ and $0.9$ in our analysis, which are motivated from the range of $\gamma$ values obtained from simulations as well as observations \citep{vogt2005,dolag,lagana}. \par
 The main results of this paper are as follows:
\begin{itemize}
\item Under the HSE condition, the average  percentage change in the value of $f_{gas,HSE}$ for $B_{0}=5\mu$G and $\gamma = 0.9$ is $0.13~\%$ at $r_{2500,HSE}$ and $0.07~\%$ at $r_{\tiny{500,HSE}}$. Similarly, using the NFW profile as a total mass estimator to gauge differences in the gas mass fraction estimates, we found that the average percentage change in $f_{gas,NFW}$ for $B_{0}= 5\mu G$ and $\gamma= 0.9$ is equal to $0.34~\%$ and $0.27~\%$ at $r_{2500,NFW}$ and $r_{500,NFW}$, respectively. These values further increase for $\gamma = 0.5$ as listed in Table~\ref{table1}.

\item Even though recent results from simulations~\citep{vazz,domi} suggest that magnetic field strengths exceeding $5\mu$G are unlikely, we have analysed the change in $f_{gas}$ for $B_{0} = 10\mu$G not only for the sake of completeness but also in the context of very precise observational programs like SKA in the coming future. As the value of magnetic field increases to $B_{0} = 10 \mu$G, the average percentage change in the value of $f_{gas,HSE}$ for $\gamma = 0.9$  increases to $0.54~\%$ and $0.31~\%$ at $r_{2500,HSE}$ and $r_{500,HSE}$, respectively. Similarly, for $\gamma = 0.5$, the average percentage change in $f_{gas,HSE}$ increases to $2.21~\%$ and $1.65~\%$ at $r_{2500,HSE}$ and $r_{500,HSE}$, respectively. Using the NFW profile as the total mass estimator provides different results similar to the case of $B_{0} = 5\mu$G as mentioned in Table~\ref{table1}.


\item The change in $f_{gas}$ is expected to increase with the strength of magnetic field and with the decrease in the value of $\gamma$ (see Section \ref{MF1}). In this regard, we have chosen a typical cluster from our sample to depict the variation of the percentage change in $f_{gas,HSE}$ with $B_{0}$ and $\gamma$ at $r_{\tiny{500,HSE}}$. This is represented in the contour plots shown in Fig. \ref{img3}, where the variation of the gas mass fraction is depicted with respect to $B_{0}$ and $\gamma$ for the cluster (CLJ0016+1609).
\end{itemize}
In summary, we have quantified the contribution of magnetic field in the gas mass fraction of galaxy clusters and found the maximum change to be $\sim 1~\%$ for $B_{0}=5\mu$G taking into account both the HSE and NFW cases and for the values of  $\gamma$ considered in our analysis. In addition, if the central magnetic field, $B_{0} \approx  10\mu$G is considered, the maximum change in $f_{gas}$ is slightly improved to $\approx 4~\%$. Although, recent results from simulations suggest that such high central magnetic field in galaxy clusters is unlikely, there still is a case for studying the consequences of such strong magnetic fields. This is because direct observations are still not precise enough but are likely to improve with new facilities in the near future. The effect of magnetic field on gas mass fraction has been discussed qualitatively in earlier studies \citep{landry} and in this work, we have attempted to quantify the same. The decrease in the gas mass fraction can have consequences on the estimates of the angular diameter distances which in turn can affect the estimates of cosmological parameters. As is evident from Fig.~\ref{img3}, the change in the gas mass fraction is dependent on the central strength of magnetic field, $B_{0}$, and the value of $\gamma$, being higher for higher values of $B_{0}$ and for lower values of $\gamma$. Hence, the effect propagated to the cosmological parameters will also in turn get enhanced with an increase in $B_{0}$ and a decrease in $\gamma$. In addition, by including the effect of other non-thermal pressures such as turbulent pressure and pressure due to cosmic rays, the gas mass fraction can further decrease leading to a much larger change in the observable cosmological parameters. 

\section*{Acknowledgements}

We thank the anonymous referee for useful suggestions and comments. We also thank Biman Nath and Kandaswamy Subramanian for useful comments. SJ, SM and TRS acknowledge the facilities at ICARD, University of Delhi. DJ acknowledges R.F.L. Holanda for useful discussions. The research of SJ is supported by INSPIRE Fellowship(IF160769), DST India and the research of SM is supported by UGC, Govt. of India under the UGC-JRF scheme (Sr.No. 2061651305 Ref.No: 19/06/2016(I) EU-V). TRS acknowledges the project grant from SERB, Govt. of India (EMR/2016/002286).

\bibliographystyle{elsarticle-num}
\bibliography{reference}

\end{document}